\documentclass[12pt,a4paper]{article}
\usepackage[utf8]{inputenc}
\usepackage[T1]{fontenc}
\usepackage[margin=1in]{geometry}
\usepackage{setspace}
\usepackage{graphicx}
\usepackage{booktabs}
\usepackage{array}
\usepackage{amsmath}
\usepackage{amssymb}
\usepackage{threeparttable}
\usepackage[hypcap=false]{caption}
\usepackage{hyperref}
\usepackage{float}
\usepackage{microtype}
\usepackage[style=apa,natbib=true,backend=biber]{biblatex}
\graphicspath{{figures/}}
\title{Asymmetric Discourse Homogenization on Reddit\\ Was a Text-Length Artifact: A Reanalysis}
\author{Fengming Liu\\ \small University of Chicago \quad \texttt{fengmingliu@uchicago.edu}}
\date{Version 2, November 2026. This version replaces version 1 (August 2026).}

\begin{document}
\maketitle

\begin{abstract}
\noindent In the first version of this paper I reported that conservative Reddit users began writing more alike after ChatGPT was released while progressive users did not. That result came from a similarity measure that rises and falls with comment length. I redo every analysis of the original paper on all 6,056,593 comments with a sentence encoder and a dispersion measure computed within length strata, and the conservative break goes away. The two camps show no reliable difference in how within-group similarity changed after late 2022, and the ChatGPT date behaves like any of the more than two thousand other cutoffs I test. The matched flair comparison, the cumulative AI index, the stayer panel and the entropy signature each lose their support once length is held fixed. The one change that survives is older. Left and right comments drew closer to each other between 2019 and 2022, and that convergence was complete before ChatGPT arrived. I end with what this implies for studies that measure homogenization with embeddings.
\end{abstract}

\section{Introduction}

In August 2026 I posted a paper arguing that a shared language technology had compressed political speech unevenly \citep{liuAsymmetricV1}. Using six million comments from r/AskConservatives and r/AskALiberal, I found that right-flaired users in the conservative forum broke from a pre-existing trend toward diversity at the ChatGPT launch, and that left-flaired users did not. The asymmetry survived interrupted time series, difference-in-differences, regression discontinuity in time, flair-level propensity-score matching and a cumulative exposure index. I read it as an ecological effect of AI on the marketplace of ideas.

The finding was an artifact. The measure, mean pairwise cosine similarity of mean-pooled DeBERTa-v3 embeddings, sits near 0.92 for random pairs of comments and rises mechanically with text length. Right-flaired comments in r/AskConservatives fell from about 109 words in 2019 to about 61 words in 2022. The ``diversification trend'' was the length decline and the ``break'' was its flattening. A companion methods paper documents the artifact in general form \citep{liuLengthArtifact2026}. Knowing that a measure failed does not tell the reader what the data show once the measure is fixed.

This version answers that question. I keep the corpus, the groups, the treatment date and every estimator from v1, and I change only the outcome. The new outcome is mean pairwise cosine similarity among comments of the same length quintile, computed from all-mpnet-base-v2 sentence embeddings of the entire corpus and averaged over the five quintiles. For unit vectors this quantity is one minus half the mean squared distance between pairs, so it does not depend on where the origin of the embedding space sits, and the stratification holds the length mix fixed from month to month.

The result is a null, and a fairly tight one. Over the original 2019 to 2025 window the corrected level shift for right-flaired users in r/AskConservatives is 0.13 pre-period standard deviations, with a 95\% interval whose upper end, 0.48, sits at half the standardized effect the original measure produces (0.96). The right-minus-left difference-in-differences is 0.04 standard deviations. When I start the series in 2021, both groups in the conservative forum show positive level shifts, and the left-flaired users show the larger one. Whatever the shorter baseline picks up belongs to the forum and carries no ideological pattern.

Every secondary result in v1 fails on the corrected measure. The failures share causes that are available to anyone who measures discourse homogenization with embeddings, so I collect them as methodological implications at the end. Section \ref{sec:ledger} lists what v1 claimed. Section \ref{sec:measure} explains the measure. Section \ref{sec:results} redoes each analysis. Section \ref{sec:camps} reports the one change that survives, a convergence between camps that predates ChatGPT. Section \ref{sec:lessons} collects the methodological implications.

\section{What the First Version Claimed}
\label{sec:ledger}

Table \ref{tab:ledger} lists each empirical claim in v1 next to its fate in this reanalysis. I state the claims as v1 stated them, so the reader can check the correspondence against the archived version.

\begin{table}[H]
\centering
\caption{Claims in v1 and their status after reanalysis}
\label{tab:ledger}
\small
\begin{tabular}{>{\raggedright\arraybackslash}p{0.42\textwidth}>{\raggedright\arraybackslash}p{0.50\textwidth}}
\toprule
Claim in v1 & Reanalysis on the corrected measure \\
\midrule
Right users in r/AskConservatives break from a diversification trend at December 2022 & Level shift 0.13 SD, $p=.45$; the 95\% interval tops out at half the standardized v1 effect \\
Left users show no break & Unchanged; the left estimate is also null \\
Asymmetry holds in DiD with group trends & Right minus left 0.04 SD, $p=.98$ \\
Asymmetry holds at monthly, weekly, daily RDiT & Null at monthly, biweekly, weekly; daily full-bandwidth jumps appear in three groups alike and vanish at a one-year bandwidth \\
ChatGPT date sits at the 50th percentile of 2,377 daily cutoffs & Now the 60th percentile for the asymmetry \\
PSM-DiD across flairs, $d=-1.11$, $p=.004$ & $d=0.00$, $p=.996$ \\
Social Conservative and Nationalist flairs move most & Still positive, and the right--left flair gap disappears once flair pre-trends are controlled \\
Cumulative LLM index survives a quadratic trend & $p=.14$; 22\% of randomly dated indices fit as well \\
Ambient detector predicts next-month author homogenization & The v1 outcome had no author-level variation; the redone test is null with month-clustered errors \\
Stayers show no change, so the effect is ecological & Stayers show no change, which says nothing about mechanism once the aggregate effect is gone; v1 measured stayers with a different embedding \\
Embeddings are \texttt{[CLS]} tokens, which encode document-level semantics & The headline series used mean pooling, and the cited source does not concern \texttt{[CLS]} \\
Word-entropy SD contracts twice as much on the right & The contraction is length dispersion; within length strata it is equal across camps \\
\bottomrule
\end{tabular}
\end{table}

\section{Data and Measurement}
\label{sec:measure}

\subsection{Corpus}

The corpus is the one used in v1. It contains 6,056,593 comments posted to r/AskConservatives and r/AskALiberal between January 2019 and December 2025, retrieved from the Arctic Shift archive \citep{heitmannArcticShift2024}, written by 37,231 authors. Ideology comes from the user flair attached to each comment. The four groups are AskConservatives/Right (1,664,712 comments), AskConservatives/Left (998,070), AskALiberal/Left (2,982,380) and AskALiberal/Right (411,431). Following v1, the treatment date is December 1, 2022, and the main comparison is AskConservatives/Right against AskALiberal/Left.

\subsection{Why the original measure failed}

v1 embedded each comment with mean-pooled DeBERTa-v3-base truncated to 256 tokens and took, for each group and month, the mean pairwise cosine among 1,000 sampled comments. Two properties of that pipeline combine into the artifact. The embedding space is strongly anisotropic, so every comment vector shares a large common component and random pairs score about 0.92 \citep{ethayarajh2019}. Mean pooling over more tokens averages away comment-specific variation, so longer comments sit closer to the common direction and to each other \citep{zhou2025lengthcollapse}. Truncating the same 20,000 comments from 150 words to 10 words lowers their mean similarity from 0.968 to 0.904, a change eight times the effect v1 reported. A placebo series in which each comment is replaced by a random comment of the same length tracks the observed conservative series at $r=0.97$ \citep{liuLengthArtifact2026}. Over 2019 to 2025 the original monthly series correlates with log mean comment length at 0.91 for AskConservatives/Right.

v1 also described its measure wrongly. Its methods section said the analysis used \texttt{[CLS]} token embeddings and cited \citet{rodriguezEmbeddingRegressionModels2023} for the claim that such embeddings encode document-level semantics. Both statements were in error. The code behind every headline series mean-pooled the token states, and \citet{rodriguezEmbeddingRegressionModels2023} develop \textit{\`a la carte} word embeddings for regression on context, which says nothing about the \texttt{[CLS]} token of a transformer. The only v1 analysis that did use \texttt{[CLS]}, truncated at 128 tokens, was the stayer analysis, so its null and the headline effect came from two different measures.

\subsection{The corrected measure}

I re-embedded all 6,056,593 comments with all-mpnet-base-v2, a sentence encoder trained contrastively for semantic similarity \citep{reimers2019sbert,song2020mpnet}. Its random-pair cosine on this corpus is 0.11. I encode up to 256 tokens, which truncates about 2\% of comments, all in the top length quintile. I then split comments into global word-count quintiles (cut points at 15, 28, 47 and 86 words). For a cell $c$ (a group in a month, a flair in a month, a group on a day) and quintile $k$ with $n_{ck}$ unit vectors $x_i$ summing to $s_{ck}$, the exact mean pairwise cosine is
\begin{equation}
h_{ck} = \frac{s_{ck}^\top s_{ck} - n_{ck}}{n_{ck}(n_{ck}-1)},
\end{equation}
and the cell measure is $H_c = \frac{1}{5}\sum_k h_{ck}$, defined when every quintile has at least five comments. Because $h_{ck} = 1 - \tfrac{1}{2}\,\overline{\|x_i - x_j\|^2}$, the measure is a monotone transform of the mean squared distance between comments. Squared distances do not change when the embedding space is translated, so subtracting a corpus mean, which alters raw cosine, leaves this quantity alone, and no renormalisation step is needed. Equal weights across quintiles mean that a month in which comments get shorter does not change $H_c$ through composition. Every comment enters, so there is no sampling noise from drawing 1,000 comments per cell.

I compute everything from sum vectors accumulated in one streaming pass over the embedding matrix, which makes daily, weekly, flair and author-type series exact, with no resampling. Two checks run on a 915,225-comment stratified sample that I also embedded with BGE-base \citep{xiao2024cpack} and with the original DeBERTa pipeline. To keep the original estimates in view I report three outcomes throughout, namely the v1 series (labelled DeBERTa), the raw mpnet cosine without stratification and the corrected measure $H$. Effects are expressed in pre-period standard deviations of the outcome, so the three can be compared.

The corrected measure does not erase length entirely. Its monthly level still correlates with mean length at 0.60 for AskConservatives/Right because both trend down over the period. After removing a linear trend from each, the correlation is 0.05. AskConservatives/Left keeps a detrended correlation of 0.42, and controlling for mean log length within quintiles moves its level shift from 0.36 to 0.32 and the AskConservatives/Right shift from 0.13 to 0.16, so the conclusions below do not depend on it.

\subsection{Estimators}

I keep the v1 estimators. The interrupted time series regresses the monthly outcome on a trend, a post-December-2022 indicator and their interaction, with Newey-West errors at six lags \citep{bernal2017its}. The difference-in-differences stacks two groups and allows group-specific pre-trends, as the v1 reconciliation table recommended. I report two baselines, the full 2019 to 2025 window used in v1 and a window that starts in January 2021. The second matters here because 2019 and 2020 carry the steepest length decline and the most unstable forum composition, and because v1 did not report it.

\section{Results}
\label{sec:results}

\subsection{The aggregate series}

\begin{figure}[H]
\centering
\includegraphics[width=\textwidth]{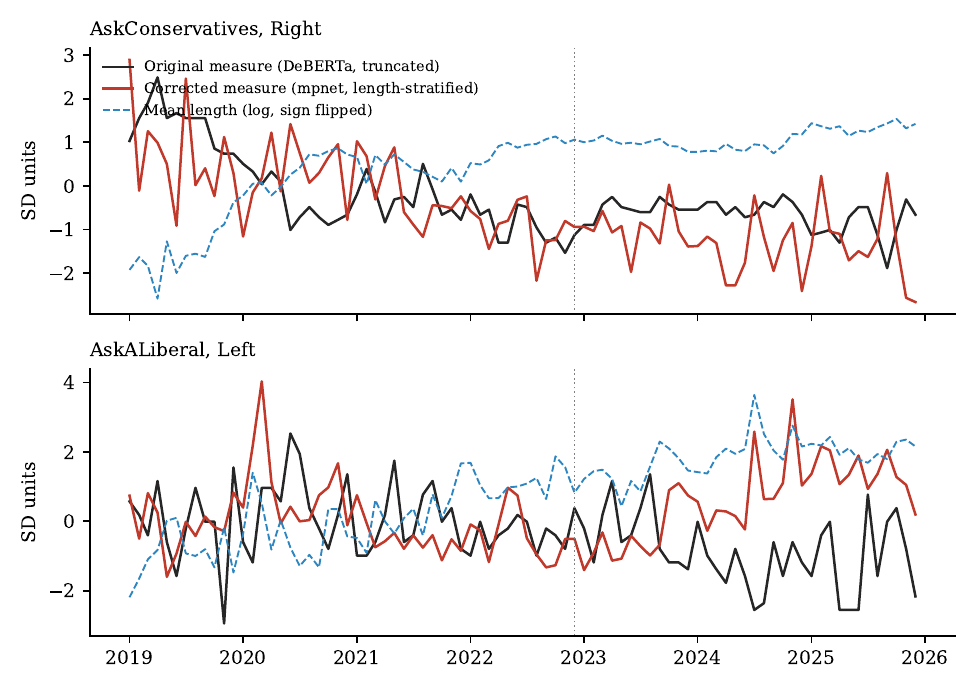}
\caption{Original and corrected within-group similarity, standardized on January 2019 to November 2022. The dashed line is log mean comment length with its sign flipped. The dotted vertical line marks December 2022.}
\label{fig:series}
\end{figure}

Figure \ref{fig:series} plots the original and corrected series for the two headline groups, standardized on the pre-period, along with mean comment length with its sign flipped. The original conservative series is a mirror image of the length series. The corrected series is noisier and shows no sustained move at December 2022.

Table \ref{tab:its} reports level shifts for all four groups. Over the v1 window, the original measure gives AskConservatives/Right a shift of 0.96 standard deviations and also gives AskALiberal/Right a shift of 1.33, a result v1 reported in a table and never discussed. On the corrected measure the AskConservatives/Right shift is 0.13 ($p=.45$, 95\% interval $-0.21$ to $0.48$). The upper bound is half the original standardized estimate. Each outcome is scaled by its own pre-period SD, so the comparison is approximate, and it still says the corrected data leave little room for an effect of the size v1 claimed. The only group with a significant corrected shift over the full window is AskConservatives/Left, at 0.36.

Starting the series in 2021 changes the estimates. Both groups in r/AskConservatives now shift up, the right by 0.61 and the left by 0.85 standard deviations. Both groups in r/AskALiberal stay null. Measured as the average gap between post-period values and the projected pre-period trend, all four groups run above trend after ChatGPT, by 2.0 to 3.2 standard deviations, and AskConservatives/Right shows the smallest gap of the four. Out-of-sample gaps of this size follow mechanically from projecting a short, declining pre-trend three years forward. Whatever the 2021 baseline picks up is common to the conservative forum and carries no ideological signature.

\begin{table}[H]
\centering
\begin{threeparttable}
\caption{Level shift at December 2022 by group, in pre-period SD}
\label{tab:its}
\footnotesize\setlength{\tabcolsep}{4pt}
\begin{tabular}{lcccccc}
\toprule
 & \multicolumn{3}{c}{Baseline 2019--2022} & \multicolumn{3}{c}{Baseline 2021--2022} \\
\cmidrule(lr){2-4}\cmidrule(lr){5-7}
Group & DeBERTa & mpnet raw & Corrected & DeBERTa & mpnet raw & Corrected \\
\midrule
AskCons./Right & $0.96^{***}$ & $0.42^{**}$ & $0.13$ & $1.42^{***}$ & $0.72^{***}$ & $0.61^{*}$ \\
 & (.000) & (.010) & (.450) & (.000) & (.001) & (.011) \\
AskCons./Left & $0.23$ & $0.42^{***}$ & $0.36^{*}$ & $0.64$ & $0.95^{***}$ & $0.85^{**}$ \\
 & (.489) & (.000) & (.012) & (.195) & (.000) & (.002) \\
AskLib./Right & $1.33^{***}$ & $0.26$ & $0.04$ & $1.46^{***}$ & $0.52$ & $0.26$ \\
 & (.000) & (.177) & (.823) & (.000) & (.119) & (.430) \\
AskLib./Left & $-0.01$ & $-0.20$ & $-0.23$ & $0.22$ & $-0.10$ & $-0.21$ \\
 & (.982) & (.565) & (.518) & (.665) & (.854) & (.713) \\
\bottomrule
\end{tabular}
\begin{tablenotes}\footnotesize
\item Coefficient on the post indicator in $y_t=\alpha+\beta t+\gamma\,\text{post}_t+\delta\, t\,\text{post}_t$, divided by the pre-period SD of $y$. $p$-values from Newey-West errors (6 lags) in parentheses. 84 months (60 for the 2021 baseline). $^{***}p<.001$, $^{**}p<.01$, $^{*}p<.05$.
\end{tablenotes}
\end{threeparttable}
\end{table}

Stratification lets me ask whether any length band carries a hidden break. None does. For AskConservatives/Right the five quintile-specific level shifts range from $-0.40$ to $0.22$ standard deviations, and none has $p<.15$.

\subsection{Right versus left}

The asymmetry was the paper's central claim, so Table \ref{tab:did} reports the difference-in-differences under each choice v1 made and each it did not. With group-specific pre-trends, the specification v1 defended, the original measure gives a right-minus-left estimate of 2.38 standard deviations ($p=.010$). The corrected measure gives 0.04 ($p=.98$). The 2021 baseline gives 1.25 ($p=.16$). Within each forum, where the comparison holds the community constant, the corrected estimates are $-0.62$ and 0.05 in r/AskConservatives and 0.25 and 0.58 in r/AskALiberal, all with $p>.44$.

Under a common trend the corrected estimates turn significantly negative, near $-2.6$ standard deviations in both windows. The groups trended at different speeds before December 2022, and forcing a shared slope assigns the gap to the post period. No specification produces a positive, significant asymmetry on the corrected measure. The sign of the asymmetry depends on which trend assumption the analyst adopts, so the data do not identify an asymmetry.

\begin{table}[H]
\centering
\begin{threeparttable}
\caption{Right minus left difference-in-differences, in pre-period SD}
\label{tab:did}
\footnotesize\setlength{\tabcolsep}{4pt}
\begin{tabular}{llcccc}
\toprule
 & & \multicolumn{2}{c}{Group-specific trends} & \multicolumn{2}{c}{Common trend} \\
\cmidrule(lr){3-4}\cmidrule(lr){5-6}
Contrast & Measure & 2019 & 2021 & 2019 & 2021 \\
\midrule
Headline (AC/R vs AL/L) & DeBERTa & $2.38^{*}$ (.010) & $0.50$ (.473) & $-0.05$ (.928) & $0.74$ (.056) \\
 & mpnet raw & $0.70$ (.603) & $1.17$ (.200) & $-2.62^{**}$ (.002) & $-2.42^{***}$ (.000) \\
 & Corrected & $0.04$ (.977) & $1.25$ (.157) & $-2.61^{***}$ (.001) & $-2.63^{***}$ (.000) \\[3pt]
Within AskCons. & DeBERTa & $1.67$ (.077) & $0.82$ (.232) & $-1.19^{*}$ (.043) & $-0.42$ (.340) \\
 & Corrected & $-0.62$ (.445) & $0.05$ (.930) & $-0.54$ (.224) & $-0.76$ (.093) \\[3pt]
Within AskLib. & DeBERTa & $1.43^{*}$ (.048) & $1.30$ (.060) & $0.29$ (.532) & $0.28$ (.472) \\
 & Corrected & $0.25$ (.865) & $0.58$ (.564) & $-0.81$ (.307) & $-0.90$ (.124) \\
\bottomrule
\end{tabular}
\begin{tablenotes}\footnotesize
\item Coefficient on Right $\times$ post, with each group's series divided by its own pre-period SD. Newey-West errors (6 lags); $p$-values in parentheses.
\end{tablenotes}
\end{threeparttable}
\end{table}

\begin{figure}[H]
\centering
\includegraphics[width=\textwidth]{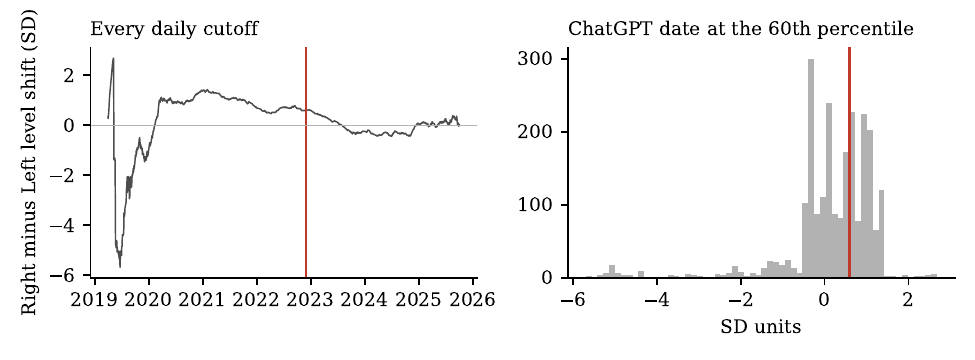}
\caption{Right minus left level shift in the corrected daily series for every cutoff from April 2019 to October 2025 (left) and its distribution (right). The red line marks December 1, 2022.}
\label{fig:perm}
\end{figure}

\subsection{Timing}

v1 already contained the evidence that the ChatGPT date was not special. Its daily permutation placed the conservative shift at the 49.8th percentile of all candidate cutoffs. v1 read that result as a warning about the date while still claiming the effect. On the corrected daily series, with every date from April 2019 to October 2025 tried as a cutoff and at least 90 days on each side, the right-wing level shift at ChatGPT ranks at the 64th percentile and the right-minus-left asymmetry at the 60th (Figure \ref{fig:perm}). The Dobbs decision ranks at the 68th percentile, the 2022 midterms at the 60th, GPT-4 at the 52nd. The most extreme cutoffs fall in June and July 2019, when the forums were small and changing fast.

The regression discontinuity estimates tell the same story with one wrinkle (Table \ref{tab:rdit}). At monthly, biweekly and weekly aggregation the AskConservatives/Right discontinuity is small and insignificant. At daily aggregation with the full bandwidth it is 0.36 standard deviations and highly significant, and so is the discontinuity for AskConservatives/Left (0.32) and AskALiberal/Right (0.34). A jump that appears in three of four groups at once cannot carry an asymmetry. It also disappears with a one-year bandwidth in every group, which identifies it as a misfit of straight lines through three years of daily data, a known weakness of RDiT with long bandwidths \citep{hausmanRegressionDiscontinuityTime2018}.

\begin{table}[H]
\centering
\begin{threeparttable}
\caption{RDiT discontinuity at December 1, 2022, corrected measure, in pre-period SD}
\label{tab:rdit}
\small
\begin{tabular}{lcccc}
\toprule
 & AskCons./Right & AskCons./Left & AskLib./Right & AskLib./Left \\
\midrule
Monthly, full & $0.13$ (.530) & $0.36^{*}$ (.026) & $0.04$ (.866) & $-0.23$ (.539) \\
Biweekly, full & $0.24$ (.207) & $0.33^{*}$ (.028) & $0.09$ (.695) & $-0.21$ (.479) \\
Weekly, full & $0.26$ (.093) & $0.34^{*}$ (.017) & $0.14$ (.451) & $-0.19$ (.419) \\
Daily, full & $0.36^{***}$ (.000) & $0.32^{***}$ (.000) & $0.34^{***}$ (.000) & $0.02$ (.812) \\
Daily, one year & $0.09$ (.525) & $0.07$ (.613) & $0.06$ (.775) & $-0.41$ (.060) \\
\bottomrule
\end{tabular}
\begin{tablenotes}\footnotesize
\item Local linear fit on each side of the cutoff, Newey-West errors with automatic lag length. ``Full'' uses all data; ``one year'' uses 365 days on each side.
\end{tablenotes}
\end{threeparttable}
\end{table}

\begin{figure}[H]
\centering
\includegraphics[width=0.9\textwidth]{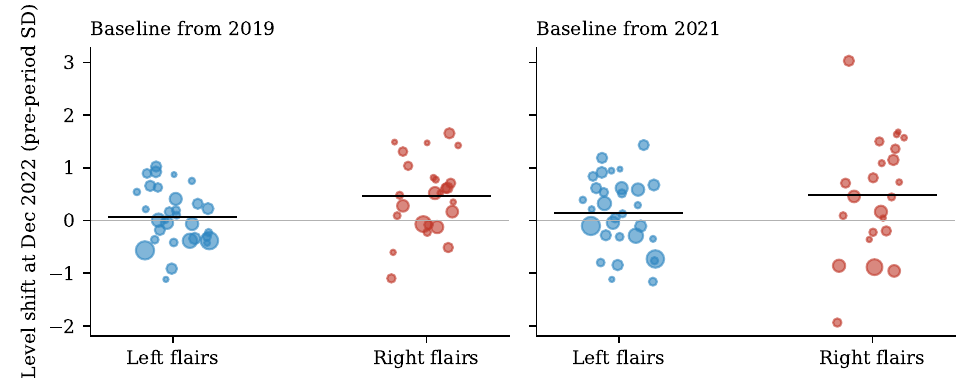}
\caption{Level shift at December 2022 for 56 flairs, corrected measure. Marker area scales with comment count; horizontal bars are unweighted means.}
\label{fig:flair}
\end{figure}

\subsection{Flairs}

v1 split the right into its flairs and found the shift concentrated in Social Conservative and Nationalist users. I rebuilt the corrected measure for the 56 flairs that meet the coverage rule on both sides of the cutoff (at least 5,000 comments overall, and at least twelve months on each side in which every length quintile has five or more comments), 25 on the right and 31 on the left. Over the full window right flairs shift up by 0.47 standard deviations on average and left flairs by 0.07, a difference of 0.37 ($p=.031$, weighted by the square root of comment counts). Social Conservative (1.31) and Nationalist (0.61) remain among the larger positive shifts (Figure \ref{fig:flair}).

The flair gap does not survive the two checks that sank the aggregate result. Right flairs had steeper pre-period declines, and a steep decline followed by a flattening is what produces a positive level shift in this model. Adding each flair's pre-period slope as a covariate in the cross-flair regression of level shifts cuts the right-minus-left difference to 0.10 ($p=.46$). Starting the series in 2021 cuts it to 0.14 ($p=.66$). Several of the flairs that move most are also among the smallest. Libertarian Conservative and Trump Supporter each contribute fewer than 5,000 comments to the analysis.

The flair-level propensity-score DiD, which v1 reported as $d=-1.11$ ($p=.004$), now returns nothing. Matching each right flair to its nearest left flair on eleven style features (comment length, words, words per sentence, unique-word ratio, type-token ratio, exclamation, question and comma rates, share of questions, word length, sentences) and on pre-period level, volatility and trend of the corrected measure, the mean difference in pre-to-post change is $-0.00001$ across 25 pairs ($p=.996$, $d=0.00$). The matching itself deserves a warning. With 31 candidate controls, matching with replacement worsened balance on nine of fourteen covariates, the pattern \citet{kingWhyPropensityScores2019} predict for propensity matching in small samples. The v1 estimate used the same design and shares this limitation, and so does the null I report here. I give the matched result little weight in either direction.

\subsection{Dose and rival covariates}

v1 replaced the post indicator with a cumulative LLM index that counts seven model releases with hand-set weights, and found that the index survived a quadratic trend when the binary indicator did not. On the corrected measure the index survives a linear trend (0.22 standard deviations per unit, $p<.001$) and fails a quadratic one ($p=.14$). The index shares 95\% of its variance with a quadratic in time, so the two cannot be separated in 84 months. I built 1,000 placebo indices by assigning the same step sizes to randomly drawn months between 2019 and 2025. Twenty-two percent of them fit the corrected conservative series at least as well as the real index. The left-wing ``placebo'' that v1 used to show asymmetry now responds more strongly to the index than the right, 0.50 standard deviations per unit under a quadratic trend ($p=.04$).

The rival covariates point the same way. Under a quadratic trend, the share of right-flaired comments mentioning Trump predicts the corrected conservative series ($p=.019$), as does the right-wing share of AskConservatives volume ($p<.001$), and adding the LLM index to either leaves it null ($p=.82$ and $p=.47$). For AskALiberal/Left, Trump mentions, Biden mentions and broad political keywords each predict the series with $p<.005$. Similarity in these forums moves with the political news cycle and with who is posting.

\subsection{Who writes}

v1 found that authors active both before and after ChatGPT showed no shift and concluded that the effect worked through the community and not through individuals. The corrected stayer series (authors with at least five comments in the group on each side of the cutoff) shows no shift either, $-0.10$ standard deviations over the full window and 0.33 from 2021 ($p=.73$ and $p=.29$). With no aggregate effect to explain, a null among stayers carries no information about mechanism. The v1 stayer null carried even less, because v1 computed it from the \texttt{[CLS]} token at 128 tokens while the aggregate series it was compared with used mean pooling at 256.

Composition did change. Authors whose first comment in a group came after December 2022 wrote 73\% of post-period comments in both r/AskConservatives groups and 42\% in AskALiberal/Left. Their comments are slightly more alike than those of stayers in the same month, by 0.005 in AskConservatives/Right, 0.006 in AskALiberal/Left and 0.014 in AskALiberal/Right ($p<.001$ each). Newcomers write more alike on both sides, so they cannot generate an asymmetry.

The author level offers a sharper test. For each author-month with at least two comments I computed the cosine between the author's mean vector and the mean vector of everyone else in the group that month, and regressed it on the post indicator with author fixed effects, a trend, mean log length and log comment count. If AI-assisted writing pulled individuals toward a shared register, typicality should rise within authors after ChatGPT. It falls slightly in three groups ($-0.06$ to $-0.09$ standard deviations, $p<.001$) and is flat in AskConservatives/Right ($-0.02$, $p=.17$). This outcome is a cosine of mean vectors and is not stratified by length, so I control for the author's mean log length and comment count in each month; the within-author design removes stable differences in how much each author writes.

\subsection{Exposure}

v1 reported that an ``ambient detector score'', the mean ChatGPT-detector probability in a subreddit-month, predicted the next-month change in an author's similarity across 62,240 author-months. In rebuilding the analysis I found that the outcome in that regression had no author-level content. The comment panel stored the group-month similarity on every comment, so each author's ``similarity'' was the group's value and each author's ``change'' was the group's monthly change, repeated once per author. The author's ``own detector score'' was likewise the group-month mean. The regression, run on 62,240 author-month cells for AskConservatives/Right, was a time series of 84 months counted many times over, an ecological inference in the sense of \citet{robinsonEcological1950}, with standard errors that ignored the repetition.

I redid it with the author typicality outcome above. With errors clustered by author the ambient score predicts next-month change in AskALiberal/Left and in both r/AskConservatives groups. Because the regressor varies only by month, the correct clustering is by month, and then every coefficient loses significance ($p=.15$ to $.87$). Collapsing to months gives a single marginal coefficient for AskALiberal/Left ($p=.049$) and nulls elsewhere. Whatever the detector score tracks, it shows no relationship with conservative convergence.

\subsection{Entropy}

v1 presented a stylistic signature to complement the embedding result. The within-month standard deviation of word-level Shannon entropy fell after 2022, by more than twice as much on the right ($d=-2.6$ against $-1.6$). The reanalysis reproduces those numbers exactly. Word entropy of a comment is close to a function of its length, and the monthly entropy SD correlates with the SD of log comment length at 0.98 on the right and 0.96 on the left. Within length quintiles the contraction is the same on both sides ($d=-1.35$ and $-1.38$). It shows no break at ChatGPT on the right ($p=.98$) and a small upward break on the left ($p=.03$), the opposite of compression. After removing a cubic in log length from each comment's entropy, the residual SD does not change at all ($d=0.10$ and $-0.14$). Character-level entropy SD, which v1 also reported, shows no pre-post change in either camp. The signature v1 described was a smaller spread of comment lengths.

\subsection{Encoders and the length placebo}

Two checks guard the corrected measure against the problems that sank the original. Table \ref{tab:enc} recomputes the corrected measure on the 915,225-comment sample under three encoders. The conservative full-window shift ranges from 0.23 to 0.38 standard deviations and no DiD reaches $p<.12$. DeBERTa, the encoder that produced the artifact, gives a null DiD once the measure is stratified by length, which locates the v1 problem in the outcome definition more than in the encoder.

The second check is the placebo that exposed v1. I replace each comment's embedding with that of a random comment of the same word count from the same group, drawn from the whole period, so the placebo series keeps every length change and loses all content. Under the raw DeBERTa measure such a placebo tracked the observed conservative series at $r=0.97$. Under the corrected measure the correlation falls to 0.17 for mpnet, 0.36 for DeBERTa and 0.41 for BGE, and no placebo DiD is significant at the 5\% level. BGE keeps some residual length sensitivity, since its placebo produces a significant shift for AskALiberal/Left from 2021, which is one reason I use mpnet for the main analysis. The corrected mpnet series carries content and little length.

\begin{table}[H]
\centering
\begin{threeparttable}
\caption{Corrected measure under three encoders and its length placebo, sample of 915,225 comments}
\label{tab:enc}
\small
\begin{tabular}{lcccc}
\toprule
 & \multicolumn{2}{c}{AskCons./Right level shift} & \multicolumn{2}{c}{Right minus left DiD} \\
\cmidrule(lr){2-3}\cmidrule(lr){4-5}
Series & 2019 & 2021 & 2019 & 2021 \\
\midrule
mpnet & $0.23$ (.221) & $0.63^{**}$ (.009) & $0.10$ (.935) & $1.29$ (.127) \\
BGE-base & $0.38$ (.090) & $0.84$ (.060) & $0.22$ (.865) & $1.01$ (.232) \\
DeBERTa (v1 pipeline) & $0.24$ (.448) & $0.60$ (.268) & $1.13$ (.228) & $0.57$ (.332) \\[3pt]
mpnet, length placebo & $-0.11$ (.688) & $-0.10$ (.819) & $-0.21$ (.764) & $-0.32$ (.646) \\
BGE, length placebo & $0.61$ (.052) & $-0.31$ (.403) & $1.52$ (.231) & $-1.03$ (.052) \\
DeBERTa, length placebo & $-0.17$ (.369) & $-0.26$ (.518) & $-0.64$ (.381) & $-0.79$ (.224) \\
\bottomrule
\end{tabular}
\begin{tablenotes}\footnotesize
\item Quintile-stratified mean pairwise cosine for each encoder. DiD allows group-specific trends. Units and errors as in Tables \ref{tab:its} and \ref{tab:did}.
\end{tablenotes}
\end{threeparttable}
\end{table}

\section{What Did Change}
\label{sec:camps}

A null on the question I asked leaves room for change on a question I did not ask. Within each forum I computed the squared distance between the left and right centroids, separately in each length quintile, from the same sum vectors. Within quintile $k$ and month this is $h_{LL}+h_{RR}-2h_{LR}$, where $h_{LR}$ is the mean cosine across camps, and I average over quintiles.

\begin{figure}[H]
\centering
\includegraphics[width=\textwidth]{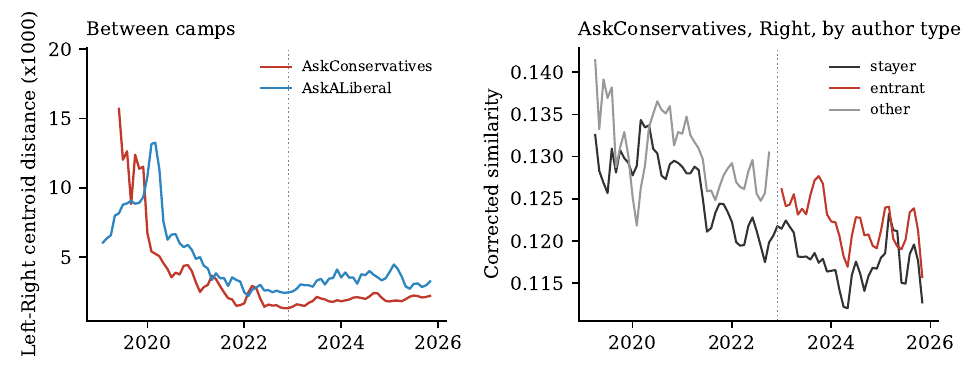}
\caption{Left panel shows the squared distance between left and right centroids within each forum, length-stratified, three-month moving average. Right panel shows corrected similarity for AskConservatives/Right by author type. ``Entrant'' authors first posted in the group after December 2022; ``other'' authors posted before but too rarely to be stayers and are shown only before December 2022, when they are numerous enough to measure.}
\label{fig:camps}
\end{figure}

The two camps converged sharply (Figure \ref{fig:camps}, left). In r/AskConservatives the distance fell from 0.0145 in 2019 to 0.0018 in 2022, and in r/AskALiberal from 0.0080 to 0.0025. Most of the decline happened by 2021. After December 2022 the distance is flat in r/AskConservatives and drifts slightly up in r/AskALiberal, and the post-ChatGPT level shift is small and positive in both forums, the opposite of convergence. Whatever brought left and right closer in these forums was in place before ChatGPT.

Because this is the one positive finding in the paper, I put it through the checks that sank v1. On the 915,225-comment sample the distance falls from 2019 to 2022 by a factor of 7.8 in r/AskConservatives and 3.6 in r/AskALiberal under mpnet, and by 6.0 and 3.5 under BGE. DeBERTa reproduces the decline in r/AskConservatives (a factor of 8.5) and shows little movement in r/AskALiberal, where its distances are an order of magnitude smaller to begin with. A length placebo, which replaces each comment's embedding with that of a random comment of the same word count from the same forum and either camp, produces distances near zero in every encoder and year, never above 0.001 in absolute value against an observed 0.0145. Length differences between camps cannot generate this series, and the decline in r/AskConservatives appears under all three encoders. I treat it as descriptive. I did not design the study to explain it, and I have no test of why the camps converged.

\section{Lessons}
\label{sec:lessons}

\textit{Measure length before measuring anything else.} Every embedding similarity measure is a function of length under some encoder and pooling choice. The cheapest diagnostic is the one that exposed v1, a placebo series that keeps each comment's length and discards its content. When the placebo reproduces the finding, the finding belongs to length. I should have run it before the first regression and I recommend it as a default for any study of homogenization in text.

\textit{Prefer quantities that cannot see the origin.} The anisotropy of contextual embeddings means raw cosine is dominated by a shared direction \citep{ethayarajh2019,steck2024cosine}. Mean squared distance within strata, equivalently stratified mean cosine of unit vectors, does not depend on that direction and does not require the renormalisation that makes centered cosine unstable. Sentence encoders reduce the problem and do not remove it. The raw mpnet series still produces a significant conservative shift in Table \ref{tab:its}.

\textit{Report the second baseline.} v1 used one window. Starting the series two years later turns a null conservative shift into a significant one and moves the largest shift to the other camp. When a result depends on where the pre-period begins, the analyst has a choice to disclose \citep{simmonsFalsePositive2011,gelmanGardenForkingPaths2014}. I now report both windows for every estimate. Multiplicity matters here as well. Applying a Holm correction to the eight group-by-window level shifts in Table \ref{tab:its} leaves one estimate below .05, AskConservatives/Left on the 2021 baseline (adjusted $p=.016$). The conservative shift on the same baseline does not survive ($p=.074$).

\textit{Let the trend assumption be visible.} The sign of most DiD estimates in this paper depends on whether groups share a pre-trend. v1 noticed this in its reconciliation table and treated it as a nuisance. When the sign flips with a modeling choice, that flip is itself the finding, and it says the data cannot separate a treatment effect from different trend speeds.

\textit{Treat the permutation test as the main test.} v1 ran a placebo over 2,377 cutoff dates, found the ChatGPT date at the median, and kept the effect. That ordering was backward. With a single treated time series and trending data, the distribution of estimates across arbitrary dates is the reference distribution, and an estimate near its median gives no reason to single out the date.

\textit{Check that individual-level outcomes vary across individuals.} The v1 exposure regression used an outcome that was constant within group-months. Its sample size, its standard errors and its interpretation as an author-level mechanism were all artifacts of a merge. A one-line check of within-cell variance would have caught it.

\textit{Dose indices built from release dates are time trends.} Any monotone step function over 2022 to 2025 correlates with time. A dose index earns evidential weight only if it beats randomly dated indices built the same way, and this one does not.

\textit{Dispersion of a length-sensitive statistic is dispersion of length.} Word entropy, type-token ratio and similar per-comment statistics depend on length. Their monthly spread tracks the spread of lengths, so a stylistic signature needs to be computed within length strata before it can corroborate anything.

\textit{Report intervals against a stated effect size.} A null from a noisy measure would have told the reader little. The corrected measure uses every comment, and its interval for the conservative shift tops out at half the standardized v1 effect. The comparison is approximate, since each outcome is scaled by its own pre-period SD, and a formal equivalence test would need a smallest effect of interest fixed in advance \citep{lakensEquivalence2018}. Even this informal comparison tells the reader more about magnitude than a bare failure to reject. The same calculation gives the design's power. With the observed standard error, the full-window ITS for AskConservatives/Right has 80\% power to detect a level shift of 0.49 pre-period standard deviations, about half the effect v1 reported, and the right-minus-left DiD with group trends needs 3.5 standard deviations. The aggregate null is informative about effects of the claimed size. The DiD null says much less, because an asymmetry test built on two noisy series with separate trends can detect only very large differences.

\section{Conclusion}

I set out to show that a shared language technology narrowed conservative speech more than progressive speech. On a measure that does not track comment length, I find no such narrowing in six million comments, no special role for the ChatGPT date, no concentration in flairs once pre-trends are respected, no dose response that a time trend cannot match, no convergence within authors and no entropy signature beyond the spread of lengths. The convergence that did happen ran between the camps and had finished before ChatGPT arrived.

This does not show that large language models leave public discourse untouched. Detector-based studies find growing LLM-assisted writing across many genres \citep{liang2025widespread,kobak2025}, and homogenization may appear in places these two forums do not reach, or at a scale smaller than this design can detect. It shows that this corpus, measured properly, does not contain the effect I reported, and that the route to my mistaken report runs through choices that other studies also make. A measure that answers before the question is asked has stopped measuring.

\section*{Data and code}
All analyses in this version run from the scripts in the replication package (folder \path{arxiv_v3_rewrite/code}, scripts 01 to 11), which read the comment corpus and the mpnet embedding matrix and write every table and figure. The v1 scripts remain in the original replication package for comparison.

\section*{Declaration}
I used AI assistants for code review and copy-editing. All analyses, interpretations and errors are mine. The error documented here was mine as well, and I thank the readers whose questions led me to it.

\printbibliography

@software{heitmannArcticShift2024,
  title = {Arctic Shift: {{Reddit}} Data Archive},
  author = {Heitmann, Arthur},
  date = {2024},
  url = {https://arctic-shift.photon-reddit.com},
  urldate = {2025-12-31},
  version = {online service}
}

@article{kingWhyPropensityScores2019,
  title = {Why {{Propensity Scores Should Not Be Used}} for {{Matching}}},
  author = {King, Gary and Nielsen, Richard},
  date = {2019},
  journaltitle = {Political Analysis},
  volume = {27},
  number = {4},
  pages = {435--454},
  doi = {10.1017/pan.2019.11},
}

@article{hausmanRegressionDiscontinuityTime2018,
  title = {Regression {{Discontinuity}} in {{Time}}: {{Considerations}} for {{Empirical Applications}}},
  shorttitle = {Regression {{Discontinuity}} in {{Time}}},
  author = {Hausman, Catherine and Rapson, David S.},
  date = {2018-10-05},
  journaltitle = {Annual Review of Resource Economics},
  shortjournal = {Annu. Rev. Resour. Econ.},
  volume = {10},
  number = {1},
  pages = {533--552},
  issn = {1941-1340, 1941-1359},
  doi = {10.1146/annurev-resource-121517-033306},
  url = {https://www.annualreviews.org/doi/10.1146/annurev-resource-121517-033306},
  urldate = {2026-02-08},
  langid = {english}
}

@article{rodriguezEmbeddingRegressionModels2023,
  title = {Embedding {{Regression}}: {{Models}} for {{Context-Specific Description}} and {{Inference}}},
  shorttitle = {Embedding {{Regression}}},
  author = {Rodriguez, Pedro L. and Spirling, Arthur and Stewart, Brandon M.},
  date = {2023-11},
  journaltitle = {American Political Science Review},
  shortjournal = {Am Polit Sci Rev},
  volume = {117},
  number = {4},
  pages = {1255--1274},
  issn = {0003-0554, 1537-5943},
  doi = {10.1017/S0003055422001228},
  url = {https://www.cambridge.org/core/product/identifier/S0003055422001228/type/journal_article},
  urldate = {2026-02-08},
  langid = {english}
}

@inproceedings{ethayarajh2019,
  author={Ethayarajh, Kawin}, title={How Contextual are Contextualized Word Representations? {Comparing} the Geometry of {BERT}, {ELMo}, and {GPT-2} Embeddings},
  booktitle={Proceedings of EMNLP-IJCNLP 2019}, pages={55--65}, year={2019}, doi={10.18653/v1/D19-1006}}

@inproceedings{reimers2019sbert,
  author={Reimers, Nils and Gurevych, Iryna}, title={Sentence-{BERT}: Sentence Embeddings using {Siamese} {BERT}-Networks},
  booktitle={Proceedings of EMNLP-IJCNLP 2019}, pages={3980--3990}, year={2019}, doi={10.18653/v1/D19-1410}}

@inproceedings{song2020mpnet,
  author={Song, Kaitao and Tan, Xu and Qin, Tao and Lu, Jianfeng and Liu, Tie-Yan}, title={{MPNet}: Masked and Permuted Pre-training for Language Understanding},
  booktitle={Advances in Neural Information Processing Systems 33}, year={2020}}

@inproceedings{xiao2024cpack,
  author={Xiao, Shitao and Liu, Zheng and Zhang, Peitian and Muennighoff, Niklas and Lian, Defu and Nie, Jian-Yun}, title={{C-Pack}: Packed Resources for General {Chinese} Embeddings},
  booktitle={Proceedings of SIGIR 2024}, pages={641--649}, year={2024}, doi={10.1145/3626772.3657878}}

@inproceedings{steck2024cosine,
  author={Steck, Harald and Ekanadham, Chaitanya and Kallus, Nathan}, title={Is Cosine-Similarity of Embeddings Really About Similarity?},
  booktitle={Companion Proceedings of the ACM Web Conference 2024}, pages={887--890}, year={2024}, doi={10.1145/3589335.3651526}}

@inproceedings{zhou2025lengthcollapse,
  author={Zhou, Yuqi and Dai, Sunhao and Cao, Zhanshuo and Zhang, Xiao and Xu, Jun}, title={Length-Induced Embedding Collapse in {PLM}-based Models},
  booktitle={Proceedings of ACL 2025 (Volume 1: Long Papers)}, pages={28767--28791}, year={2025}, doi={10.18653/v1/2025.acl-long.1396}}

@article{bernal2017its,
  author={Lopez Bernal, James and Cummins, Steven and Gasparrini, Antonio}, title={Interrupted Time Series Regression for the Evaluation of Public Health Interventions: A Tutorial},
  journal={International Journal of Epidemiology}, volume={46}, number={1}, pages={348--355}, year={2017}, doi={10.1093/ije/dyw098}}

@article{liang2025widespread,
  author={Liang, Weixin and Zhang, Yaohui and Codreanu, Mihai and Wang, Jiayu and Cao, Hancheng and Zou, James}, title={The Widespread Adoption of Large Language Model-Assisted Writing across Society},
  journal={Patterns}, volume={6}, number={12}, pages={101366}, year={2025}, doi={10.1016/j.patter.2025.101366}}

@article{kobak2025,
  author={Kobak, Dmitry and Gonz{\'a}lez-M{\'a}rquez, Rita and Horv{\'a}t, Em{\H{o}}ke-{\'A}gnes and Lause, Jan}, title={Delving into {LLM}-Assisted Writing in Biomedical Publications through Excess Vocabulary},
  journal={Science Advances}, volume={11}, number={27}, pages={eadt3813}, year={2025}, doi={10.1126/sciadv.adt3813}}

@online{liuLengthArtifact2026,
  author={Liu, Fengming}, title={Embedding Cosine Measures Text Length Before It Measures Homogenization},
  date={2026}, eprinttype={arxiv}, eprint={2611.21519}, eprintclass={cs.CL}}

@online{liuAsymmetricV1,
  author={Liu, Fengming}, title={Asymmetric Discourse Homogenization and Shared Language Technology: Evidence from {Reddit}},
  date={2026}, eprinttype={arxiv}, eprint={2608.13674}, version={1}, note={Version 1}}

@article{simmonsFalsePositive2011,
  author={Simmons, Joseph P. and Nelson, Leif D. and Simonsohn, Uri}, title={False-Positive Psychology: Undisclosed Flexibility in Data Collection and Analysis Allows Presenting Anything as Significant},
  journal={Psychological Science}, volume={22}, number={11}, pages={1359--1366}, year={2011}, doi={10.1177/0956797611417632}}

@article{gelmanGardenForkingPaths2014,
  author={Gelman, Andrew and Loken, Eric}, title={The Statistical Crisis in Science},
  journal={American Scientist}, volume={102}, number={6}, pages={460--465}, year={2014}, doi={10.1511/2014.111.460}}

@article{lakensEquivalence2018,
  author={Lakens, Dani{\"e}l and Scheel, Anne M. and Isager, Peder M.}, title={Equivalence Testing for Psychological Research: A Tutorial},
  journal={Advances in Methods and Practices in Psychological Science}, volume={1}, number={2}, pages={259--269}, year={2018}, doi={10.1177/2515245918770963}}

@article{robinsonEcological1950,
  author={Robinson, W. S.}, title={Ecological Correlations and the Behavior of Individuals},
  journal={American Sociological Review}, volume={15}, number={3}, pages={351--357}, year={1950}, doi={10.2307/2087176}}

\end{document}